# Simultaneous Nitrogen-Doping and Reduction of Graphene Oxide

Xiaolin Li, Hailiang Wang, Joshua T. Robinson, Hernan Sanchez, Georgi Diankov, Hongjie Dai*

*Department of Chemistry, Stanford University, Stanford, California 94305*

*E-mail: hdai@stanford.edu*

Abstract

We develop a simple chemical method to obtain bulk quantities of N-doped, reduced graphene oxide (GO) sheets through thermal annealing of GO in ammonia. X-ray photoelectron spectroscopy (XPS) study of GO sheets annealed at various reaction temperatures reveals that N-doping occurs at a temperature as low as 300ºC, while the highest doping level of ~5% N is achieved at 500ºC. N-doping is accompanied by the reduction of GO with decreases in oxygen levels from ~28% in as-made GO down to ~2% in 1100ºC $NH_3$ reacted GO. XPS analysis of the N binding configurations of doped GO finds pyridinic N in the doped samples, with increased quaternary N (N that replaced the carbon atoms in the graphene plane) in GO annealed at higher temperatures (>900ºC). Oxygen groups in GO were found responsible for reactions with $NH_3$ and C-N bond formation. Pre-reduced GO with fewer oxygen groups by thermal annealing in $H_2$ exhibits greatly reduced reactivity with $NH_3$ and lower N-doping level.

Electrical measurements of individual GO sheet devices demonstrate that GO annealed in $NH_3$ exhibits higher conductivity than those annealed in $H_2$, suggesting more effective reduction of GO by annealing in $NH_3$ than in $H_2$, consistent with XPS data. The N-doped reduced GO shows clearly n-type electron doping behavior with Dirac point (DP) at negative gate voltages in three terminal devices. Our method could lead to the synthesis of bulk amounts of N-doped, reduced GO sheets useful for various practical applications.

## Introduction

Graphene exhibits various interesting physical properties,1 large surface areas (~2600 $m^2/g$) and high chemical stability, all of which could be utilized for potential applications including graphene nanoribbon field effect transistors,2,3 graphene sheet supercapacitors 4 and lithium secondary batteries.5 Chemical doping is important to modulate the electrical properties of graphene. Devising doping methods for this two dimensional material will be key to its future applications and require drastically different approaches from conventional methods for bulk materials.6-11 Recently, we reported N-doping of individual graphene nanoribbons through electrical joule heating in $NH_3$ and suggested reactions occurring mostly at the edges and defect sites on graphene.6 Substitutional N-doped multiplayer graphene sheets were synthesized by Liu et al. by adding $NH_3$ gas during chemical vapor deposition (CVD) growth of graphene.11 N-doped graphite was also prepared by arc discharge of carbon electrodes in the presence of $H_2$/pyridine or $H_2$/ammonia.12 Currently, systematic investigations of graphene doping are still needed.

Here we use GO as a starting material to investigate the reaction with $NH_3$ at elevated temperatures. We investigate the N-doping and reduction effect of annealing of GO in $NH_3$ by XPS characterization and electrical measurements, in side-by-side comparisons to GO annealed in $H_2$. The results lead to insights to the degree of doping and reduction effects at various temperatures, the roles played by oxygen groups at graphene edges and defect sites in the reaction and electrical properties of the resulting graphene sheets. Our method also provides an effective method for the synthesis of gram-scale N-doped reduced GO sheets, which could lead to useful properties unattainable by un-doped graphene.

## Results and discussion

Graphene oxide (GO) sheets were synthesized from graphite powder using a modified Hummers method (detail synthesis procedure in supplementary information).13,14 Due to harsh oxidization, GO sheets have limited sizes (from several hundred nanometers to 1 or 2 microns, Fig. 1a) and disrupted conjugation in the plane. There are vacancies of carbon atoms in the plane, and abundant functional groups such as epoxide, hydroxyl, phenol, carbonyl, carboxyl, lactone, and quinone at both the edges and defects in the plane (Fig. 1a).15-21 We used atomic force microscopy (AFM) to characterize GO sheets deposited on the substrate from the supernatant and observed that GO was mostly single-layer sheets of various shapes and sizes with the apparent thickness of about 1nm (Fig. 1a), corresponding to single layer GO.15

Reaction with ammonia was done by annealing GO samples in a 2 torr

$NH_3$/Ar (10% $NH_3$) atmosphere. For sample preparation, as-made GO sheets were deposited on $SiO_2$ substrates from solution and dried to form thick films or lyophilized to obtain fluffy powders for reactions and subsequent XPS characterization. We carried out GO annealing in $NH_3$ with the samples heated in a $NH_3$ flow from room temperature to various temperatures up to 1100ºC (detailed doping process in supplementary information). Control experiments were done by annealing GO samples in 2 torr $H_2$ at the same temperatures. We used XPS to characterize the elemental composition of GO sheets reacted under various conditions (Fig. 1b and 1c). As-made GO sheets showed more than ~28% oxygen and no nitrogen signal in the XPS spectrum (Fig. 1b, c and 2). The high-resolution C1s XPS spectrum of as-made GO sheets showed a second peak at higher binding energies (Fig. 3a and 3b), corresponding to large amounts of $sp^3$ carbon with C-O bonds, carbonyls (C=O) and carboxylates (O-C=O),15-21 resulted from harsh oxidation and destruction of the $sp^2$ atomic structure of graphene (Fig. 1a).

XPS revealed that N-doping occurred at a temperature as low as 300°C for GO annealed in $NH_3$, with ~3.2% N detected in the sample (Fig. 1b, 2a).

N levels in GO sheets annealed in $NH_3$ between 300 and 1100°C were in the range of ~3-5%, with 500°C annealing affording the highest N-doping level of ~5% (Fig. 2a). In addition to N-doping, $NH_3$ annealing of GO also showed obvious reduction effect. Comparing the oxygen levels of GO samples annealed in $NH_3$ and $H_2$ at various temperatures, we found that the oxygen levels in GO annealed in $NH_3$ were lower than in those annealed in $H_2$ at the same temperatures except for 1100°C (Fig. 2b). This indicated more effective reduction effects of thermal annealing in $NH_3$ than in $H_2$ below ~1100°C.

GO samples annealed in $NH_3$ and $H_2$ both showed much lower signals at the higher binding energy end of the C1s peak than as-made GO, indicating thermal annealing in $NH_3$ and $H_2$ removed functional groups and $sp^3$ carbon (Fig. 3a and 3b). Detailed analysis of the full width of half maximum of the C1s peak at 284.5eV (graphite-like $sp^2$ C) showed that GO samples annealed in $NH_3$ exhibited wider C1s peaks than those annealed in $H_2$ (Fig. 3c). This was likely due to N incorporation into the $sp^2$ network of reduced GO samples upon annealing in $NH_3$. It is known that, rather than a single symmetry peak with a constant width, the C1s peak of $sp^2$ carbon at 284.5eV becomes asymmetric and broadened towards the high binding energy side as the amount of functional groups increases.15-21 GO annealing in $NH_3$ led to ~3 to 5% N incorporation into the sheets to afford C-N bonded groups. Thus, GO annealed in $NH_3$ exhibited broader C1s peaks than GO annealed in $H_2$ (Fig. 3c), due to N-doping and C-N species.15,21-23

For GO exposed to 2 torr of $NH_3$/Ar (10% $NH_3$) at room temperature, XPS revealed no N signal after pumping the sample in vacuum. The clear N signals in our high temperature N-doped samples were not due to physisorbed $NH_3$ but covalent C-N species formed during $NH_3$ annealing. We also dispersed N-doped GO samples in various solvents like dichloroethane, alcohols, and $H_2O$ by sonication and then deposited the GO on substrates for further XPS analysis. XPS showed no change in N-doping level before and after sonication, suggesting formation of covalent C-N bonds in the samples instead of physisorbed $NH_3$ (XPS data not shown). We also tried to further anneal N-doped GO sheets (made by 700°C annealing in $NH_3$) in $H_2$ up to 900°C and found the N levels in the samples were stable. We observed no significant decrease of the N level up to 900°C annealing in $H_2$ (Fig. S2). Heat treatment of N-doped graphite in the range of 900 to 1200°C is necessary to break C-N bonds and remove nitrogen.21

We investigated the bonding configurations of N atoms in the $NH_3$ annealed GO sheets based on high-resolution N1s XPS spectra. The N1s peaks in the XPS spectra of GO annealed at 300°C to 1100°C were fitted into two peaks, a lower energy peak A and a higher energy peak B (Fig. 4a-c and Fig. S5). In all the samples, peak A is near 398.3eV, corresponding to pyridinic N (Fig. 4d).11,21-23 The binding energy of the high energy peak B increased with annealing temperature (Fig. 4a-c and Fig. S5), indicating different N bonding configurations in GO sheets reacted with $NH_3$ at different temperatures. In GO annealed at 300-500°C, the N bonding configurations of component B could be indexed to amide, amine or pyrrolic N.11,21-23 For samples annealed at high temperatures, i.e. ≥ ~900°C, the peak position of component B was near 401.1eV (Fig. 4d), which could be indexed to quaternary N, i.e. N that replaced the carbon atom in the graphene sheets and bonded to three carbon atoms (Fig. 4d inset)11,21-23. Our data suggested that higher temperature annealing of GO in $NH_3$ above ~900°C afforded more quaternary N incorporated into the carbon network of graphene. Raman spectra were used to characterize the N-doped GO. The G peak position of GO annealed in $NH_3$ at 1100°C showed an obvious downshift (Fig. S6).

To glean the reaction pathway between $NH_3$ and GO, we carried out control experiments by performing 900°C $NH_3$ annealing of pre-reduced GO sheets made by annealing in $H_2$ at various temperatures ranging from 300°C to 1100°C. Detectable amount of N was only observed by XPS in GO pre-reduced in $H_2$ below ~500°C (Fig. S1). The N levels were below the XPS detection limit in GO samples reduced in $H_2$ at higher temperatures (>500°C) and then reacted with $NH_3$ at 900°C (Fig. S1). These experimental findings suggested that certain oxygen functional groups in the as-made GO were responsible for reactions with $NH_3$ to form C-N bonds and afford N doping. These oxygen functional groups were mostly removed from GO by heating to ~500°C, resulting in much reduced reactivity with $NH_3$ and little

N-doping in graphene. It is known in graphite oxide that carboxylic and lactone groups begin to decompose at about 250°C and –COOH and carbonyl groups are reduced by 450°C heat treatment.21 Phenol and quinone groups decompose almost entirely between 500 and 900°C. Above 900°C, –OH groups start to decrease and oxygen-containing groups in graphite oxide are completely eliminated at ~1100°C.21 Taken together, we suggest that oxygen functional groups in GO including carbonyl, carboxylic, lactone, and quinone groups are responsible for reacting with $NH_3$ to form C-N bonds. For GO reduced in $H_2$ by annealing above ~500°C, these reactive oxygen groups are decomposed and removed, thus leading to much reduced reactivity with $NH_3$ seen in our experiments.

For higher quality and lower defect density graphene sheets24 and graphene nanoribbons2 produced by mild oxidization, we expect lower reactivity with $NH_3$ at elevated temperatures and thus lower N-doping levels than GO. Oxygen groups existing at the edges and defect sites in the plane of high quality graphene could react with $NH_3$ in a similar manner as in GO, giving rise to N-doping effects.6 In our current work, we used GO as a model system to investigate the reaction with $NH_3$, since GO contained large numbers of functional groups at defect and edge sites, which gave sufficiently high N doping levels easily detected by spectroscopy.

To investigate how N-doping affects the electronic properties of graphene, we made single-sheet, back-gated electrical devices of GO (with Pd source-drain S-D contacts, Fig. 5a) after annealing in $NH_3$ and $H_2$ at high temperatures (500-900°C) (Fig. 5 and Fig. S3, S4). Figure 5 showed typical electrical devices of GO annealed in $NH_3$ and $H_2$ respectively at 700°C and 900°C. GO annealed in $NH_3$ and $H_2$ both showed p-type behavior in air, which was due to doping by physisorbed molecular oxygen and polymers involved in device fabrication.6 To avoid the complication from physisorbed oxygen, we measured the devices in vacuum (~$5x10^{-6}$ Torr). The Dirac point (DP) of the GO sheet annealed in $NH_3$ at 500°C (Fig. S3), 700°C (Fig.5b) and 900°C (Fig.5c) was at negative gate voltages of $V_{gs}$ < ~-20V in vacuum, an indication of n-type electron doping behavior due to N-dopants in graphene. After high-bias electrical annealing6 to further remove physisorbed species, the $NH_3$-annealed GO sheet showed completely n-type behavior with DP moved to highly negative gate voltages (Fig. 5b, 5c). Lower resistance was observed for GO annealed at 900°C than those annealed at 700°C due to more effective reduction of GO at higher temperatures. The Dirac point (DP) of the GO sheet annealed in $NH_3$ at 900°C was also at negative gate voltages of $V_{gs}$ < ~ -20V in vacuum (Fig. 5c). In contrast, the DP positions of $H_2$ annealed GO sheet were near the intrinsic $V_{gs}$ = 0V (Fig. 5d) and remained so after electrical annealing (Fig. 5d). The results confirmed n-doping by N dopants in graphene afforded by thermal reaction with $NH_3$.

Electrical measurements confirmed that thermal annealing of GO in $NH_3$ was effective for GO reduction in agreement with XPS data. The normalized resistance (defined as $R \cdot W/L$, where $R$ is the measured resistance of the graphene device, $W$ and $L$ are the graphene sheet width and channel length respectively) of GO reduced in $NH_3$ at 500 to 900°C were lower at the minimum conductivity DP point than GO annealed in $H_2$ at the corresponding temperatures (Fig. 5e). This was consistent with that thermal annealing in $NH_3$ afforded more effective reduction effect than annealing in $H_2$, as revealed by XPS (Fig. 2b). For comparison with other reduction methods, GO annealed in $NH_3$ at 900°C showed higher conductivity than the same GO reduced by a recently reported hydrazine solvothermal reduction method at 180°C,25 although the normalized resistance of the 900°C $NH_3$ annealed GO was still more than 100 times higher than that of pristine graphene due to the irreversible defects such as large vacancies and disrupted conjugation in the plane resulted from harsh oxidization.24-26

## Conclusion

In summary, we obtained up to 5% N-doped, reduced GO sheets by thermal annealing GO in $NH_3$. The chemical doping and reduction effects were elucidated by XPS characterization and electrical transport measurements. Oxygen groups such as carboxylic, carbonyl, and lactone groups were suggested to be essential for reactions between graphene and $NH_3$ for C-N bond formation. For different graphene samples with varying degrees of oxidation, the degree of reaction with ammonia and N-doping will depend on the amounts of these oxygen functional groups at the defect and edge sites of graphene. We expect that N-doped, reduced graphene sheets could be used for further functionalization chemistry and for various potential applications including in the clean energy area.

## Figure captions

**Figure 1.** Annealing of GO in $NH_3$ and $H_2$. (a) Schematic structure and AFM image of GO sheets. Left panel: Schematic structure of GO sheets. The conjugated plane is disrupted. There are missing carbon atoms in the plane, and functional groups like epoxide (1), hydroxyl (2), phenol (3), carbonyl (4), carboxyl (5), lactone (6), and quinone (7) at both the edges and in the plane.15-21 Right panel: A representative AFM image of GO sheets. (b) XPS spectra of GO sheets annealed in 2 torr of $NH_3$/Ar (10% $NH_3$) at various temperatures. (c) XPS spectra of GO sheets annealed in 2 torr $H_2$ at various temperatures.

**Figure 2.** N-doping and reduction effects of GO. (a) Nitrogen percentage in the GO sheets annealed in $NH_3$ at various temperatures detected by XPS. (b) Oxygen percentage in the GO sheets annealed in $NH_3$ and $H_2$ respectively at various temperatures detected by XPS. The error bars are based on more than 3 different spots measured over the samples.

**Figure 3.** XPS spectra of C (1s) peaks. (a) High resolution C (1s) spectra of as-made GO and GO annealed in $NH_3$ at different temperatures. (b) High resolution C (1s) spectra of as-made GO and GO annealed in $H_2$ at different temperatures. (c) The C (1s) peak widths of as-made GO and $NH_3$ and $H_2$

annealed GO sheets respectively, measured from the full width of half maximum of the peak at 284.5eV. Dashed line is the corresponding C (1s) peak width of pristine HOPG (highly oriented pyrolytic graphite) as a reference.

**Figure 4.** XPS spectra of N dopants in graphene. (a) High resolution N (1s) spectra of GO annealed in $NH_3$ at 300°C. The peak is fitted into a low and high energy A and B two components centered at 398.5eV and 399.9eV respectively. (b) High resolution N (1s) spectra of GO annealed in $NH_3$ at 700°C. The peak is fitted into a low and high energy A and B two components centered at 398.3eV and 400.8eV respectively. (c) High resolution N (1s) spectra of GO annealed in $NH_3$ at 900°C. The peak is fitted into a low and high energy A and B two components centered at 398.2eV and 401.1eV respectively. (d) Positions of peak A and B for different $NH_3$ annealing temperatures. Inset is a schematic structure showing the two predominant binding conditions of nitrogen in graphene annealed at high temperatures > 900°C.

**Figure 5.** Electrical properties of single GO sheet annealed in $NH_3$ vs $H_2$. (a) A typical AFM image of a N-doped GO sheet device. The right panel depicts the device structure with a 300nm thick $SiO_2$ as gate dielectric and heavily

doped Si substrate as back-gate. (b) Current-gate voltage ($I_{ds}$-$V_{gs}$) curves (recorded at $V_{ds}$ =1 V) of a single GO device fabricated with an $NH_3$-annealed (700°C) GO sheet. Red solid line: device measured in air. Green solid line: device measured in vacuum. Blue solid line: device measured in vacuum after electrical annealing. (c) Current-gate voltage ($I_{ds}$-$V_{gs}$) curves of a single GO device fabricated with an $NH_3$-annealed (900°C) GO sheet. Red solid line: device measured in air. Green solid line: device measured in vacuum. Blue solid line: device measured in vacuum after electrical annealing. (d) Current-gate voltage ($I_{ds}$-$V_{gs}$) curves of a single GO device fabricated with a $H_2$-annealed (900°C) GO sheet. Red solid line: device measured in air. Green solid line: device measured in vacuum. Blue solid line: device measured in vacuum after electrical annealing.6 (e) Statistics of normalized sheet resistance of devices fabricated with single GO sheets annealed in $NH_3$ and $H_2$ at different temperatures. The error bars are based on more than 20 different devices measured. Normalized resistance is defined as $R{\cdot}W/L$, where $R$ is resistance of device, $W$ and $L$ are the GS width and channel length respectively.

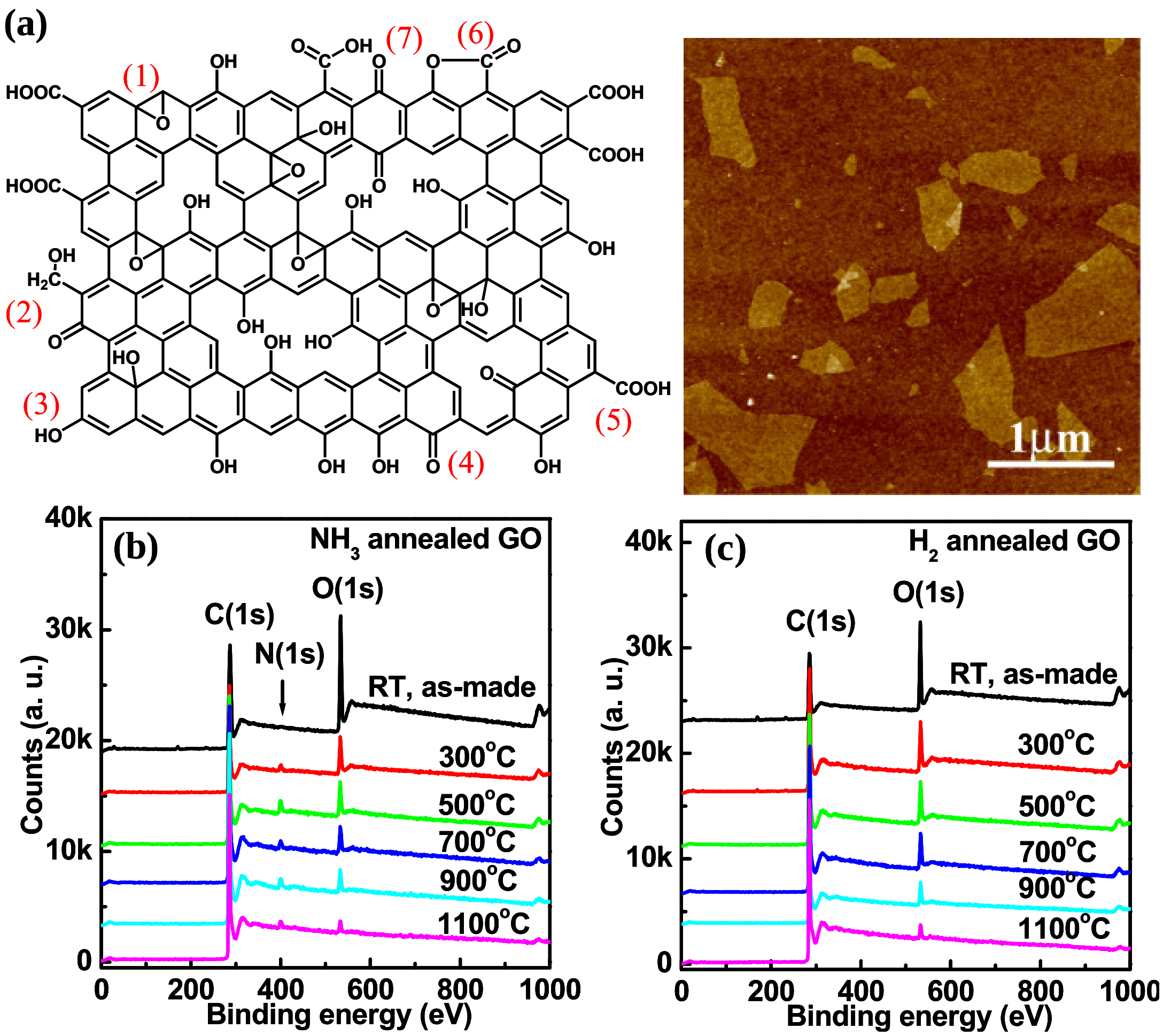

**Figure 1**

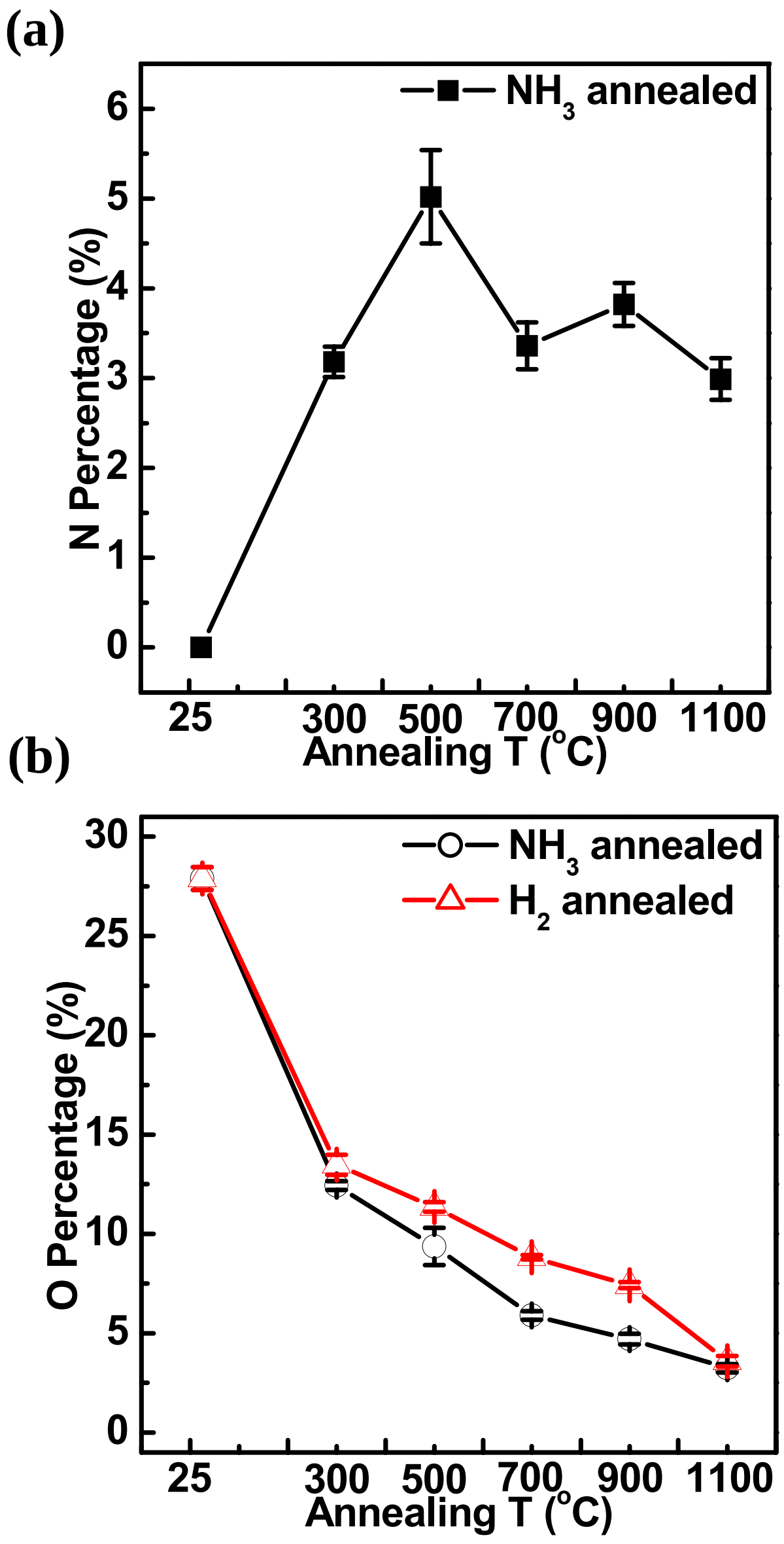

Figure 2

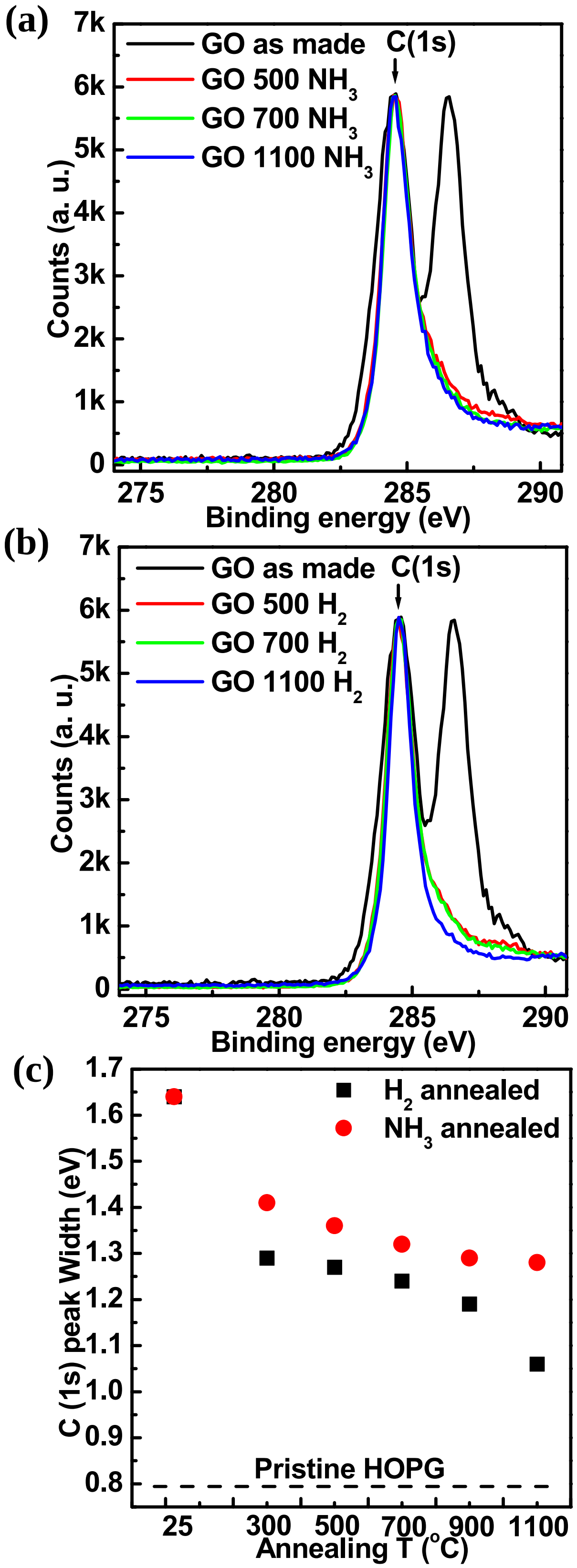

Figure 3

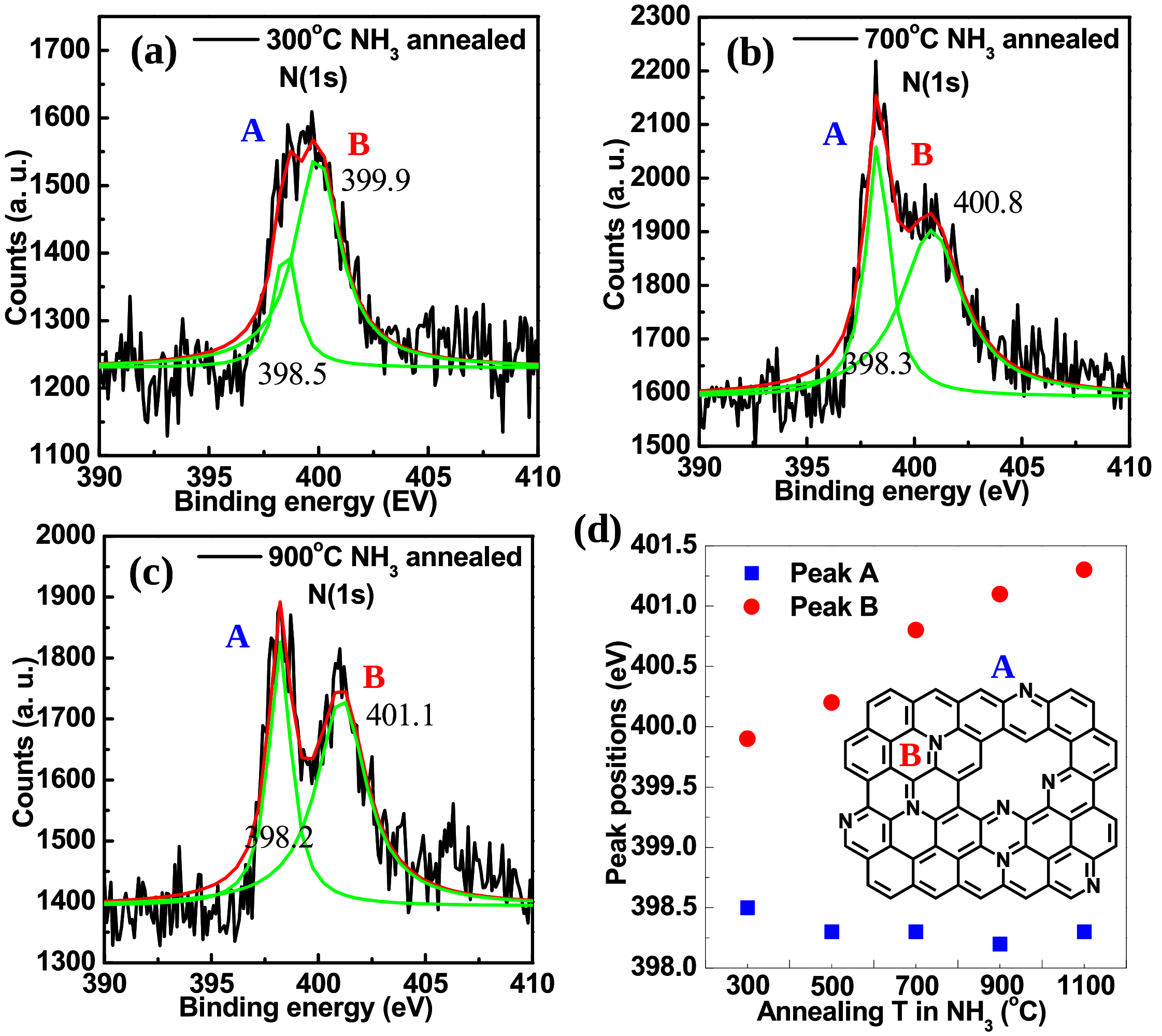

**Figure 4**

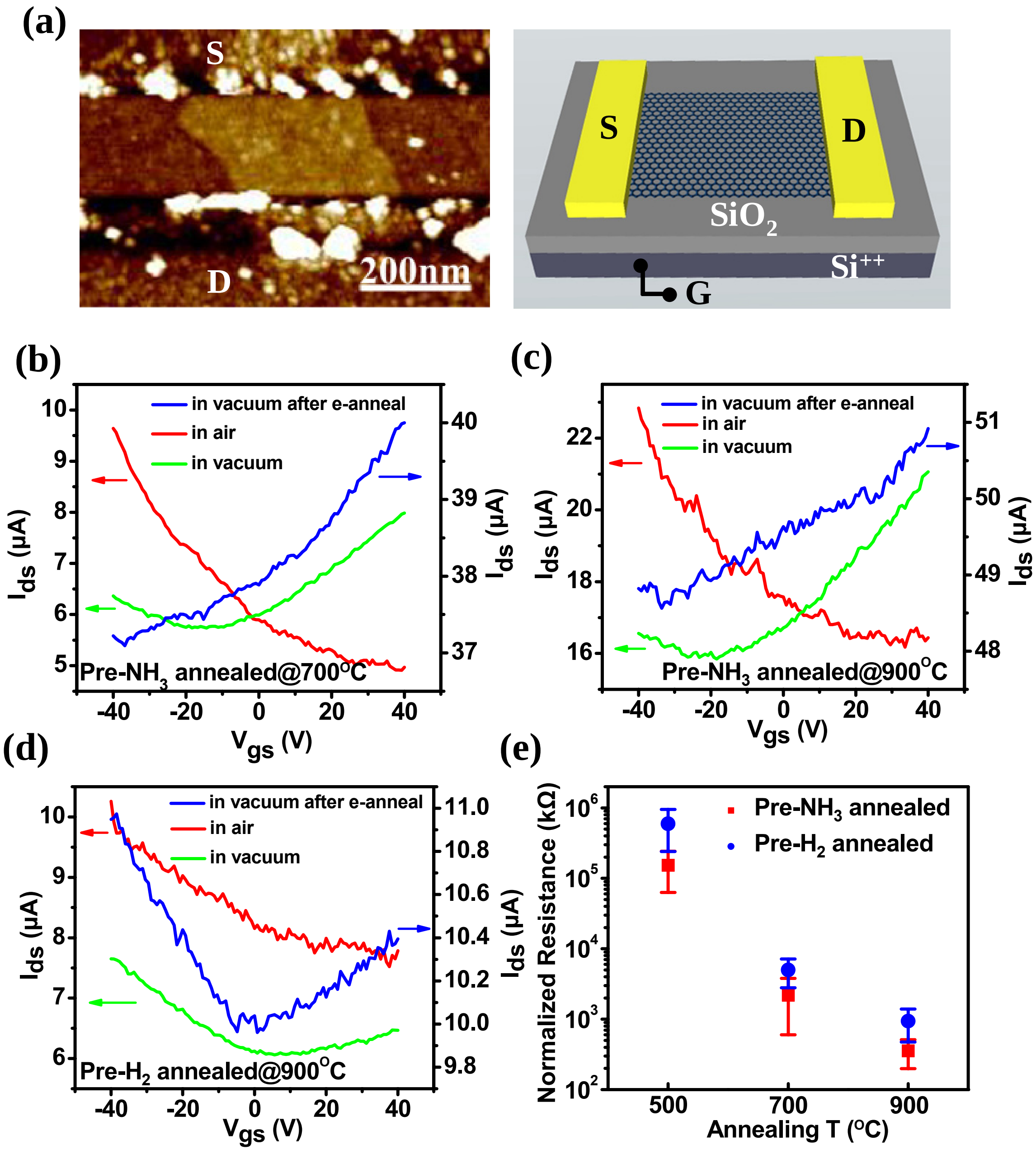

**Figure 5**